\documentclass[reprint,aps,prl]{revtex4-1}
\usepackage{epsfig}
\usepackage{graphicx}
\usepackage[tbtags]{amsmath}

\begin{document}

\title{Experimental demonstration of error-insensitive approximate universal-NOT gates}

\author{Sang Min Lee$^1$}
\author{Jeongho Bang$^2$}
\author{Heonoh Kim$^1$}
\author{Hyunseok Jeong$^2$} 
\author{Jinhyoung Lee$^3$}
\author{Han Seb Moon$^1$}
\email{moon.hanseb@gmail.com}
\affiliation{$^1$Department of Physics, Pusan National University, Busan 609-735, Korea}
\affiliation{$^2$Center for Macroscopic Quantum Control, Department of Physics and Astronomy, Seoul National University, Seoul 151-747, Korea}
\affiliation{$^3$Department of Physics, Hanyang University, Seoul 133-791, Korea}

\date{\today}

\begin{abstract}
We propose and experimentally demonstrate an approximate universal-NOT (U-NOT) operation that is robust against operational 
errors. In our proposal, the U-NOT operation is composed of stochastic unitary operations represented by the vertices of 
regular polyhedrons. The operation is designed to be robust against random operational errors by increasing the number of 
unitary operations (i.e., reference axes). Remarkably, no increase in the total number of measurements 
nor additional resources are required to perform the U-NOT operation. Our method can be applied in general to reduce
operational errors to an arbitrary degree of precision when approximating any anti-unitary operation in a stochastic manner.
\end{abstract}


\maketitle

%\newcommand{\bra}[1]{\left<#1\right|}
%\newcommand{\ket}[1]{\left|#1\right>}
%\newcommand{\abs}[1]{\left|#1\right|}
%\newcommand{\expt}[1]{\left<#1\right>}
%\newcommand{\braket}[2]{\left<{#1}|{#2}\right>}
%\newcommand{\commt}[2]{\left[{#1},{#2}\right]}
%\newcommand{\tr}[1]{\mbox{Tr}{#1}}

%section
{\it Introduction.-}
For the implementation of quantum information processing, it is necessary to reduce errors and their effects on whole processes.
Quantum information processing such as quantum computing and communications is composed of three stages: state preparation,
operations, and measurements. Any physical process that is, either implicitly or explicitly, described within the framework of quantum mechanics consists of these three stages, and errors may occur during any processes.
Errors are detrimental to the final outcome in the measurement stage: for example, both the decoherence process that
changes the state before the final measurement and the inefficiency of the measurement device (i.e., detector) will affect the
results. Recently, it was reported that the inaccuracy of unitary operations such as incorrect changes in the references for
measurements plays a crucial role in diminishing quantum effects~\cite{Jeong}. We refer to these types of errors as
``operational errors'' in contrast to the errors caused by state decoherence or inefficient detection. Schemes
including the composite pulse technique~\cite{Jones1, Jones2} and quantum error correction codes~\cite{NC} have been
suggested to reduce or correct various types of errors, but these methods require additional resources such as a larger
number of pulses~\cite{Jones2} or ancillary qubits~\cite{Shor} to enhance the precision of quantum information processing.

Anti-unitary operations in quantum mechanics are non-physical operations, and thus they cannot be implemented in a perfect
manner. However, approximate implementations are possible~\cite{MartiniN,MartiniPRA,Lim}, and some implementations such as
the universal-NOT (U-NOT) gate~\cite{MartiniN} and the transpose operation~\cite{Lim} are particularly useful for quantum
cloning, quantum state estimation, and entanglement detection~\cite{Buzek, Enk, Martini3, Gisin, Peres, Lim}. The approaches
taken for the implementation of anti-unitary operations can be categorized into two types: ancilla-assisted
models~\cite{MartiniN} and stochastic mapping~\cite{MartiniPRA, Lim}, as outlined in \cite{Bang}. The necessary condition for
the universality of each approach has been shown to be two ancillary qubits or three stochastic operations.

There are two important factors for implementing a U-NOT gate. One factor is the average value of the fidelities to the\
target state for all possible input states, and the other is their standard deviation, known as the
``universality''~\cite{Bang}. Typically, the average fidelity is the dominant factor when estimating the accuracy of a
quantum operation. However, the fidelity deviation may also be important in some cases~\cite{Magesan}. An approximate U-NOT gate is one such 
example~\cite{MartiniN} in which the fidelity is 2/3 for any input state so that the zero fidelity deviation is always 
guaranteed. In fact, there may be situations in which the fidelity deviation is practically important. For example, certain 
tasks such as fault-tolerant quantum computing~\cite{FTQC} may require the fidelity to be above a certain limit. In addition, 
supposing that an ensemble of a pure state is the input for a quantum operation but we do not know which pure state it is in 
(the input is an unknown state), then it may be important to reduce the fidelity deviation, i.e., the sensitivity of the
final results to the input state or operational errors.

In this paper, we propose and experimentally demonstrate a method to effectively reduce the effects of operational errors on 
the operation of an approximate U-NOT gate. The U-NOT gate is designed to be insensitive to operational errors by increasing 
the number of reference axes without increasing any resources or the total number of measurements. The experiment in this 
study was performed for stochastic mapping, but the same method is applicable to a certain ancilla-assisted model. We used 
spontaneous parametric down-conversion (SPDC) and linear optics elements in the experimental realization, and the stochastic 
map was characterized by quantum process tomography (QPT)~\cite{NC}. From the results of the QPT, we calculated the 
sensitivities of the maps, which matched well with the simulations and analytic predictions. In principle, our method to 
reduce the effects of operational errors can be applied to any type of approximate anti-unitary operation realized in a 
stochastic manner.

%section
{\it Concept \& theory.-} The U-NOT gate is represented by the mapping $\left| \psi \right \rangle \mapsto \left| \psi_\bot \right \rangle$, where 
$\left| \psi \right \rangle$ is an {\em arbitrary unknown} input state in a qubit, and $\left| \psi_\bot \right \rangle$ is its orthogonal state. It is well 
known that such a gate cannot be completely realized but only approximately implemented \cite{MartiniN}. To evaluate the 
approximate U-NOT gate, we introduce two measures, the average fidelity $F$ and fidelity deviation $\Delta$, defined as
\begin{eqnarray}
F = \int f(\psi)  d\psi,~~\Delta = \sqrt{\int f(\psi)^2 d\psi - F^2},
\end{eqnarray}
where $f(\psi)$ is the fidelity between the orthogonal state and the output state of the approximate operation $O$ for a pure 
input state $\left| \psi \right \rangle$, i.e., $f(\psi)=\left \langle \psi_\bot | O(\psi) | \psi_\bot \right \rangle$. Note that $F$ can be 
maximized to $2/3$ (the so-called optimality condition) and $\Delta$ can be $0$ (the so-called universality condition); these
are regarded as the best conditions for realizing an optimum approximate U-NOT gate.

In Ref.~\cite{Bang}, it was demonstrated that the approximate U-NOT gate can be realized with {\em more than or equal to} three 
stochastic unitary operations such that $\rho \mapsto O(\rho) = \sum_{i=1}^m p_i U_i \rho U_i ^\dagger ~ (m \ge 3)$, 
where $U_i$ is a single-qubit unitary operation given by $U_i$$=$$\cos{\frac{\theta_i}{2}} I + 
\sin{\frac{\theta_i}{2}}\left(\vec{\sigma}\cdot{\vec{n}_i}\right)$. Here, $\vec{\sigma}=(\sigma_1, \sigma_2, \sigma_3)$ 
is a vector operator whose elements are the Pauli operators, $\vec{n}_i=(n_{i1}, n_{i2}, n_{i3})$ is a normalized (real) 
vector, i.e., $|\vec{n}_i|=1$, and $\{p_i\}$ is the probability distribution of the stochastic operations such that 
$\sum p_i=1$. 
The necessary condition for obtaining the maximum $F$ is that the rotation angles $\{\theta_i\}$ are $\pi$ (for all 
$i=1,2,\ldots,m$). However, perfect universality (i.e., $\Delta=0$) is achieved by choosing an 
appropriate set of normalized directional vectors $\{ \vec n _i \}$ and a suitable probability distribution $\{p_i\}$.
We find conditions of $\{\vec n _i\}$ for a uniform distribution of $\{p_i\}$: $\{\vec n _i\}$ point to the vertices of
regular polyhedrons~\cite{poly} that are equally distributed in solid angle. Based on the above descriptions, our generalized 
stochastic process for the approximate U-NOT operation is as follows:
\begin{eqnarray}
\rho \rightarrow \rho_{N} '=O_{N} (\rho)=\frac{1}{N}\sum_{i=1}^{N} \left(\vec{\sigma}\cdot{\vec{n}_i}\right) \rho 
\left(\vec{\sigma}\cdot{\vec{n}_i}\right).
\label{eq_poly} 
\end{eqnarray}

For the cases of $O_3$ and $O_4$, the $\{ \vec n _i \}$ are given by $\{$(1,0,0), (0,1,0), (0,0,1)$\}$ and 
$\{(\frac{1}{\sqrt{3}}, \frac{1}{\sqrt{3}}, \frac{1}{\sqrt{3}})$, $(\frac{1}{\sqrt{3}}, \frac{-1}{\sqrt{3}} , 
\frac{-1}{\sqrt{3}})$, $(\frac{-1}{\sqrt{3}}, \frac{1}{\sqrt{3}}, \frac{-1}{\sqrt{3}})$, $(\frac{-1}{\sqrt{3}}, 
\frac{-1}{\sqrt{3}}, \frac{1}{\sqrt{3}})\}$, which correspond to the vertices of an octahedron and tetrahedron, respectively.
We can easily generalize these to the cases of $O_{6}$ and $O_{8}$ by considering the opposite directional vectors 
$\{- {\vec n} _i \}$. If error-free (in the ideal case), all the maps $O_N$ in Eq.~(\ref{eq_poly}) are equivalent to that of $N$ = 3, the Hillery-Bu\v{z}ek U-NOT gate~\cite{Buzek2, Barnett} as  $\rho \mapsto O_I(\rho)=\frac{1}{3} \left( \sigma_x \rho \sigma_x + \sigma_y \rho \sigma_y + \sigma_z \rho \sigma_z  \right)$. Thus, Eq.~(\ref{eq_poly}) is the optimum approximate U-NOT gate.

We now consider the maps $O_N$ with errors by taking realistic circumstances into account. Errors usually deteriorate 
the average fidelity and fidelity deviation in implementations of the approximate U-NOT gate, and here, we consider a 
specific but very common (operational) error that arises from the imperfect setting of $U_i$. It is 
important to note that $\Delta$ can be seriously affected by even a small error, whereas 
$F$ will remain close to its maximum value of $2/3$~\cite{Bang}. This trend motivated us to invent 
an {\em error-insensitive} approximate U-NOT gate, significantly reducing the influence of the errors on both of $F$ and $\Delta$.

We show that {\em adding more stochastic operations will increase the resilience against the operational errors}.
This can be verified by analytic calculations for the cases of $N$ = 3, 4, 6, and 8. The process in 
Eq.~(\ref{eq_poly}) is characterized by $\chi$-matrices through $O(\rho)=\sum_{i,j=0}^{3} \chi_{ij}  \sigma_i  \rho  
\sigma_j$, where $\sigma_0 = I$~\cite{NC}, and so the ideal case of the approximate U-NOT gate is characterized
by $\chi_I=\textrm{diag}(0, \frac 1 3, \frac 1 3, \frac 1 3)$. However, operational errors occurring under realistic 
circumstances will vary $\chi_I$. By using the $\chi$-matrix, we can find the mean of average fidelity $\overline{F}_N$ and fidelity deviation $\overline{\Delta}_N$ 
for the map $O_N$ over the random errors as
\begin{eqnarray}
\overline{F}_{N} \simeq \frac{2}{3}, ~~ \overline{ \Delta}_{N} \simeq \frac{\alpha}{\sqrt{N}}\delta_r=\mathcal{S}_N \delta_r,
\label{dev}
\end{eqnarray}
where $\alpha$ is a constant factor, and $\delta_r$ is the standard deviation of the random error. Since $\overline{F}_{N}$ is stationary, we define the 
error sensitivity of $O_N$ with $\overline{ \Delta}_{N}$ as $\mathcal{S}_N \equiv \alpha / \sqrt{N}$. In deriving Eq.~(\ref{dev}), the random error is 
considered as a random unitary operation $V_i$ following the stochastic operation, e.g., $\vec \sigma \cdot \vec n 
_i$$\xrightarrow{{error}}$$V_i (\vec \sigma \cdot \vec n _i)$. We consider the error operation as $V_i=e^{i ~  \vec \epsilon _i \cdot \vec \sigma} \simeq  I+ i ~ \vec \epsilon_i \cdot  \vec \sigma$, where $\vec \epsilon _i =(\epsilon _{i1},\epsilon _{i2},\epsilon _{i3} )$ and $|\epsilon_{ij}|<\epsilon_0 \ll 1$, so that the error distribution is {\em symmetric} and homogeneous (see \cite{SP} for more details). Eq.~(\ref{dev}) shows directly that the deterioration of $\Delta_N$ due to operational errors can be reduced by simply increasing the number of stochastic operations $N$.

Note that $N$ is the number of stochastic operations that constitute the map $O_N$ and is not the number of measurements. However, 
we see that Eq.~(\ref{dev}) is very similar in form to that for the standard error of $N$ measurements of random variables. 
The reason for this similarity is that the output state of the map is a mixed state: a convex combination of $N$ states. An 
erroneous output state of $O_N$ is expressed as a summation of vectors in the Bloch sphere as 
$\sum_{i=1}^N (\vec{o}_i + \vec e _i)/N$, where $\vec o _i$ are the Bloch vectors of ideal stochastic operations, and 
$\vec e _i$ are effects of the operational errors~\cite{f2}. The total effect of operational errors is described as 
$\sum_{i=1}^N \vec e _i/N$ with an average of zero and a standard deviation proportional to
$\Delta(|\vec e _i|)/\sqrt{N}$. Therefore, the average fidelity remains close to the 
maximum, and a mean of $\Delta_N$ for an erroneous map is expressed by Eq.~(\ref{dev}).

\begin{figure}[t]
\centerline{\includegraphics[scale=0.4]{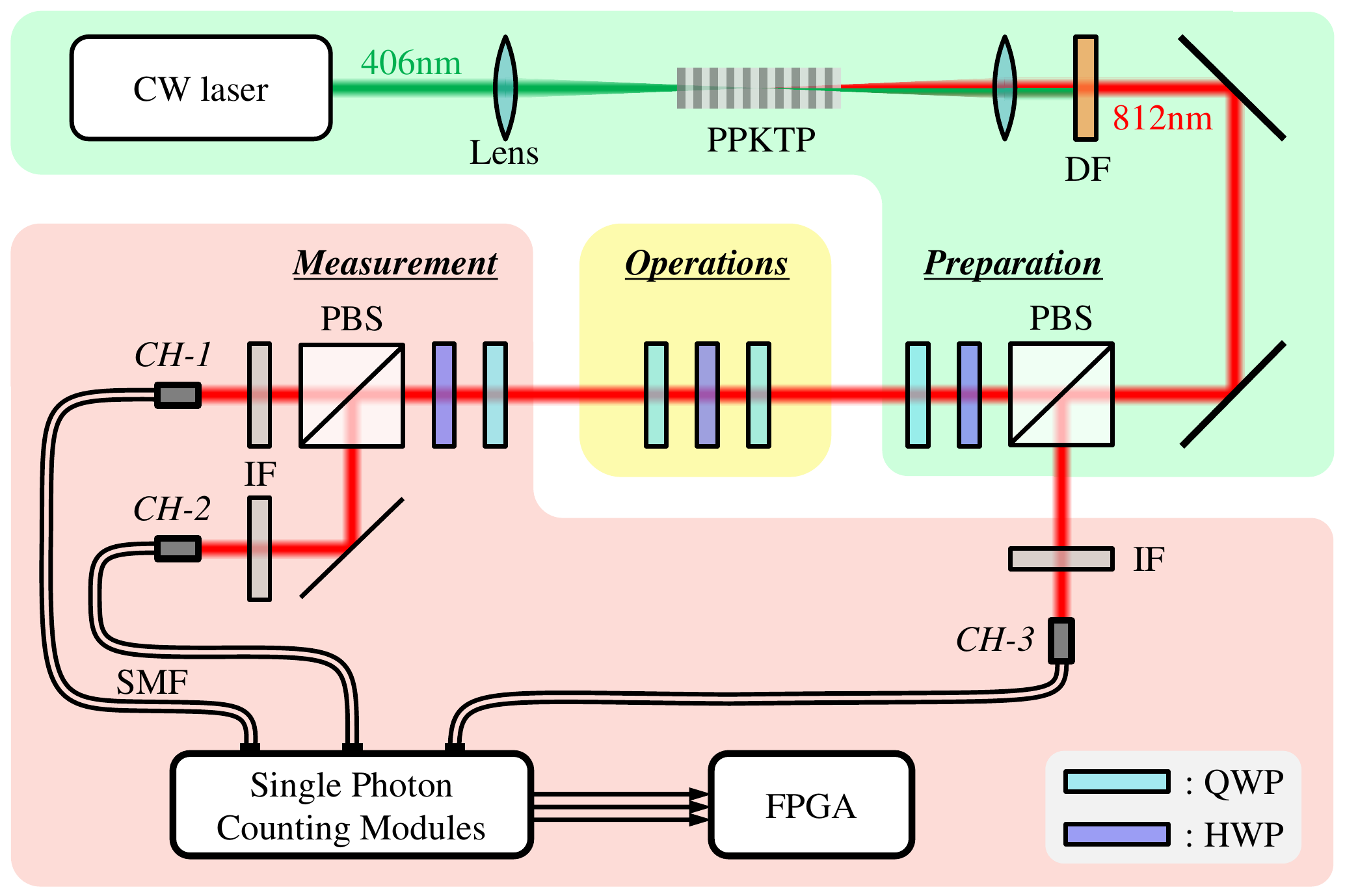}}
\caption{Experimental setup. PPKTP: periodically-poled potassium titanyl phosphate (KTiOPO$_4$), DF: dichroic filter, PBS: 
polarizing beam splitter, QWP: quarter-wave plate, HWP: half-wave plate, IF: interference filter, SMF: single-mode fiber, and 
FPGA: coincidence counter.}
\label{setup}
\end{figure}

%section
{\it Setup \& Method.-}
The experiment was based on the polarization state of a single photon generated from SPDC and manipulated by linear optics, 
as shown in Fig.~\ref{setup}. We first generated a pair of photons via a frequency-degenerate collinear type-II SPDC 
process using a diode continuous-wave laser (1.8 mW at 406 nm) and a periodically-poled KTiOPO$_4$ crystal (PPKTP; 
$L$ = 10 mm, $\Lambda$ = 10.00 $\mu$m). The polarizations of the two photons in a single mode (the same frequency and spatial mode) were orthogonal to each other as $|HV\rangle$.
A horizontal photon transmitting the polarizing beam splitter (PBS) was controlled to be in an arbitrary polarization qubit 
state through the use of the half-wave plate (HWP) and the quarter-wave plate (QWP). The vertical photon plays the role of 
a counting trigger for coincidence counts.

An arbitrary unitary operation for the polarization qubit can be realized by a set of wave plates 
(QWP--HWP--QWP)~\cite{Damask}. We stochastically perform unitary operations $\{\vec \sigma \cdot \vec n _i\}$ to realize the map $O_N$. The random error of a unitary operation is achieved by rotating each wave plate in the set 
randomly between [$-\phi_{e},\phi_{e}$].

The output states are measured by a polarization analyzer (QWP--HWP--PBS) and reconstructed by quantum state tomography 
(QST)~\cite{QST}. To obtain $F_N$, $\Delta_N$, and $\mathcal{S}_N$ of each stochastic map $O_N$, we execute QPT, which 
characterizes a quantum operation by means of the QST results for four input states and their output states. From the result of QPT and 
$\chi$-matrix, we can calculate $F_N$ and $\Delta_N$~\cite{SP}. To survey $\mathcal{S}_N$, we repeat the 
QPT measurements of $O_N$ for $N$ = 3, 4, 6, and 8 by varying the boundaries of the random error $\phi_e$ from 0$^\circ$ to 
5$^\circ$~\cite{EM}.

%figures

\begin{figure}[t]
\begin{tabular}{cccc}
\tiny{N=3}
&
\tiny{N=4}
&
\tiny{N=6}
&
\tiny{N=8}
\\
\includegraphics*[scale=0.32,viewport=59 88 229 265]{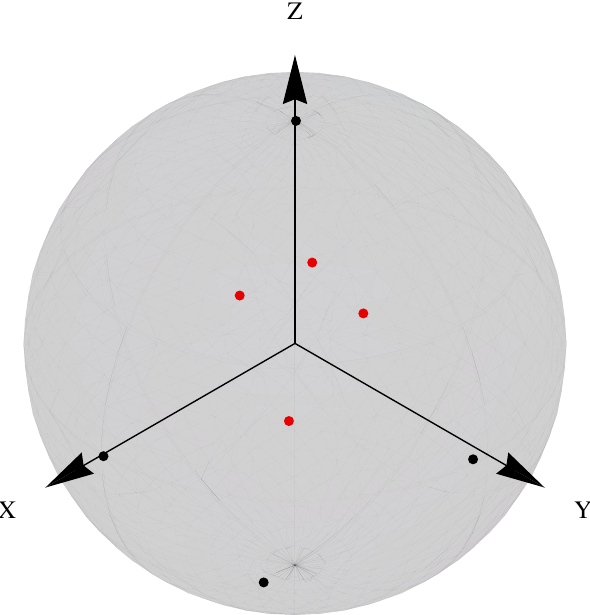}
&
\includegraphics*[scale=0.32,viewport=59 88 229 265]{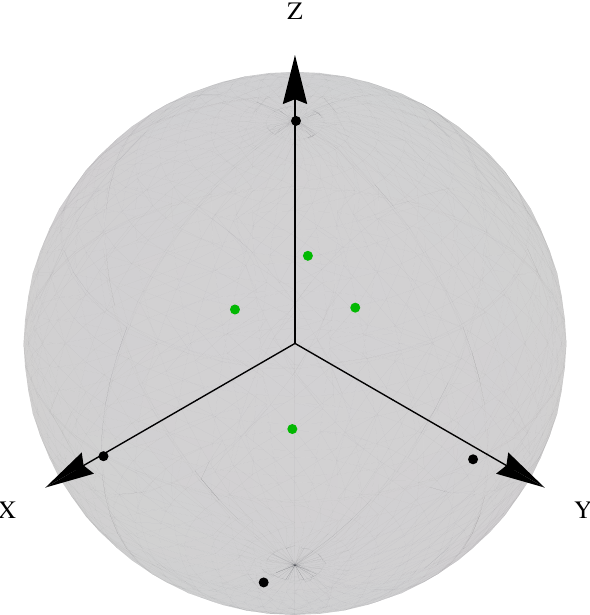}
&
\includegraphics*[scale=0.32,viewport=59 88 229 265]{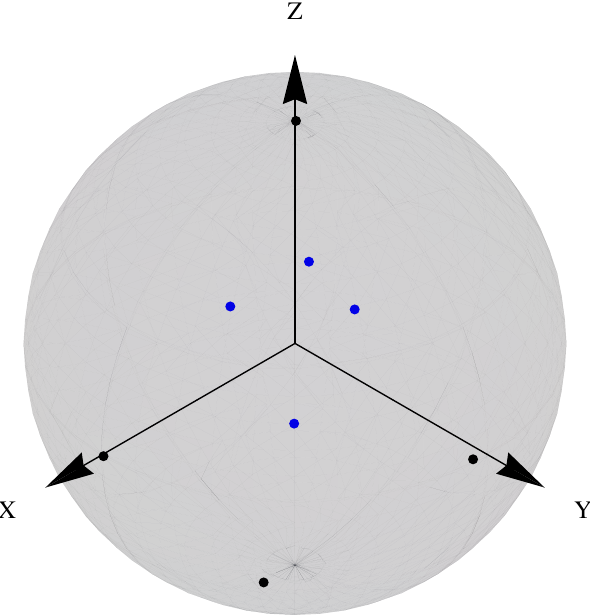}
&
\includegraphics*[scale=0.32,viewport=59 88 229 265]{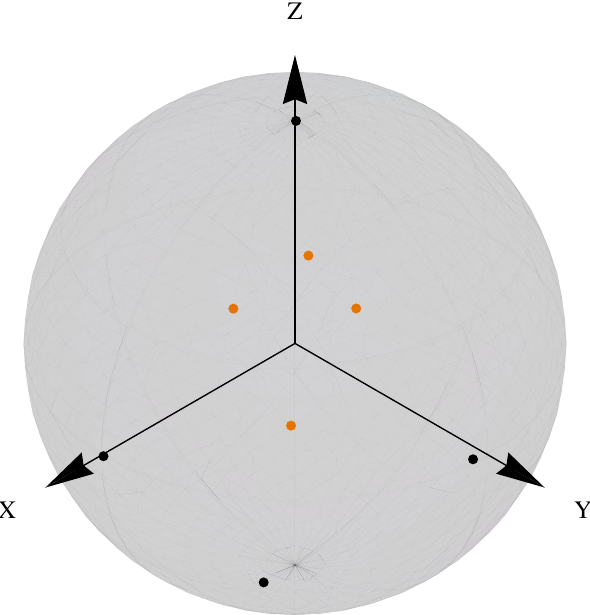}
\\
\tiny{Re[$\chi_3$]}
&
\tiny{Re[$\chi_4$]}
&
\tiny{Re[$\chi_6$]}
&
\tiny{Re[$\chi_8$]}
\\
\includegraphics*[scale=0.45]{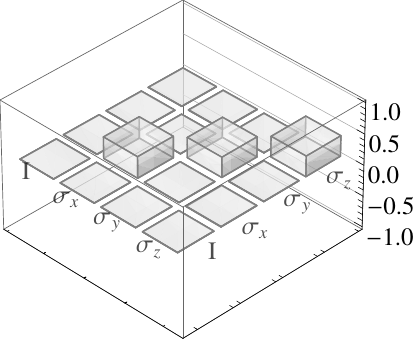}
&
\includegraphics*[scale=0.45]{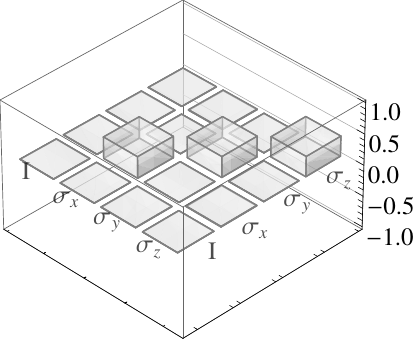}
&
\includegraphics*[scale=0.45]{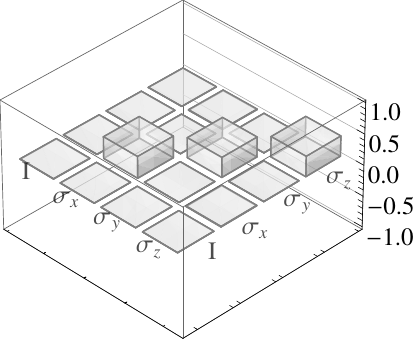}
&
\includegraphics*[scale=0.45]{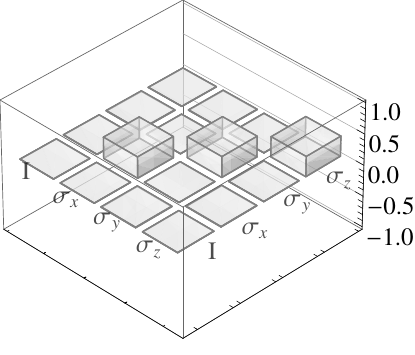}
\\
\tiny{Im[$\chi_3$]}
&
\tiny{Im[$\chi_4$]}
&
\tiny{Im[$\chi_6$]}
&
\tiny{Im[$\chi_8$]}
\\
\includegraphics*[scale=0.45]{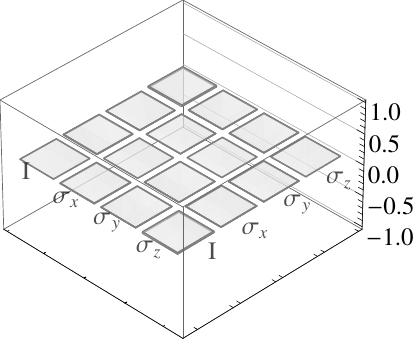}
&
\includegraphics*[scale=0.45]{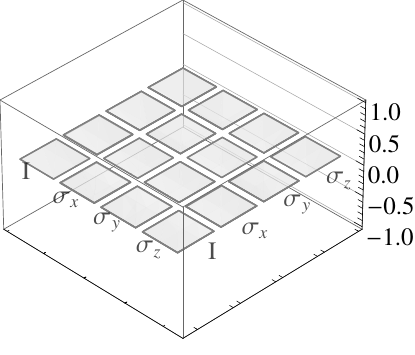}
&
\includegraphics*[scale=0.45]{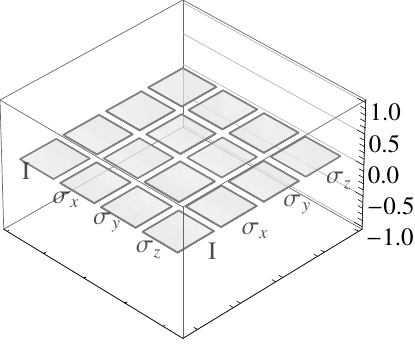}
&
\includegraphics*[scale=0.45]{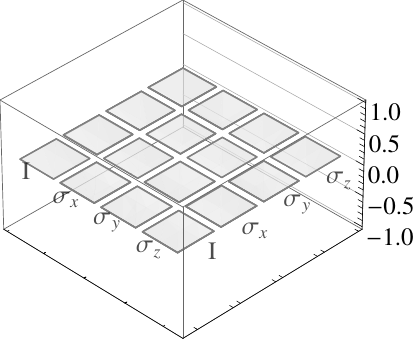}
\end{tabular}
\caption{Experimentally reconstructed input and output states of the maps $O_{3,4,6,8}$ for QPT and their $\chi$-matrices 
under error-free circumstances.}
\label{inout}
\end{figure}

%section
{\it Results \& analysis.-}
Figure \ref{inout} shows reconstructed input states and their output states for QPTs of the maps $O_{3,4,6,8}$ in the Bloch 
spheres and $\chi_N$-matrices when there are no operational errors. The points (in black) on the surface are the input states, and the 
points (in red, green, blue, and orange) close to the center represent the output states of the maps $O_{3,4,6,8}$~\cite{pcf}. The 
graphs show clearly that the output states are on the opposite sides of the input states and that their lengths decrease by 
about 1/3. The $\chi_N$-matrices are calculated from the reconstructed density matrices of the input and output states, and 
these are almost the same as those of the ideal case, i.e., $\chi_{11}$ = $\chi_{22}$ = $\chi_{33}$ = 1/3. This constitutes 
experimental verification of the equivalence of the maps $O_N$ for $N$ = 3, 4, 6, and 8 under error-free circumstances.

%figures

%figures
\begin{figure}[t]
\begin{tabular}{ccc}
\tiny{N=3}
&~
&
\tiny{N=4}
\\
\includegraphics*[scale=0.5,viewport=59 88 229 265]{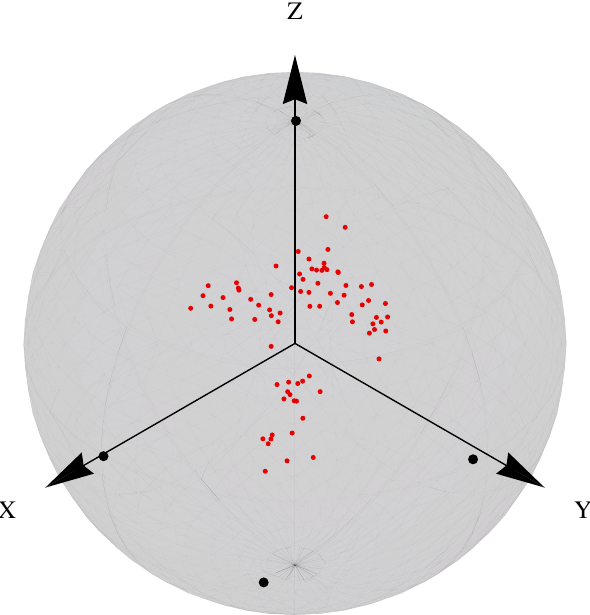}
&~
&
\includegraphics*[scale=0.5,viewport=59 88 229 265]{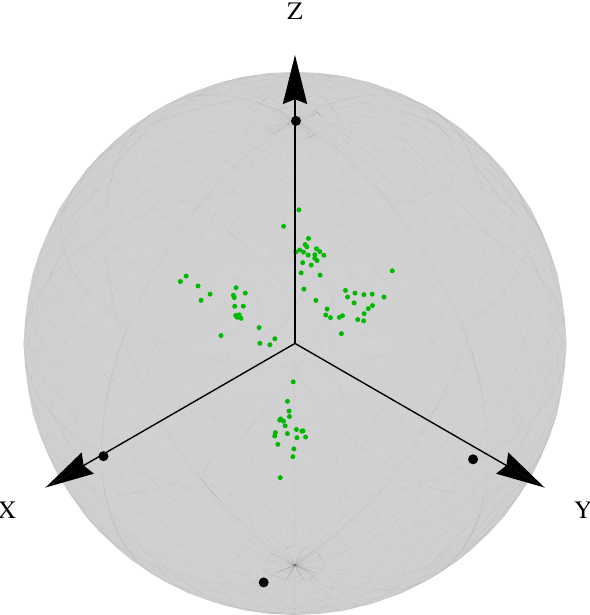}
\\
\tiny{N=6}
&
&
\tiny{N=8}
\\
\includegraphics*[scale=0.5,viewport=59 88 229 265]{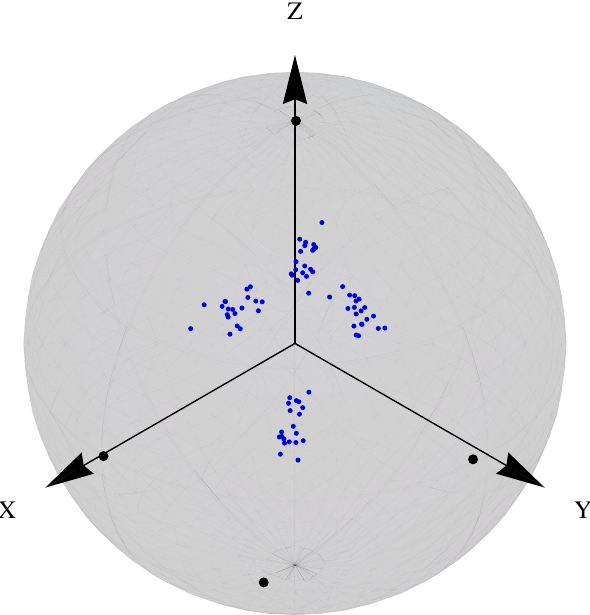}
&~
&
\includegraphics*[scale=0.5,viewport=59 88 229 265]{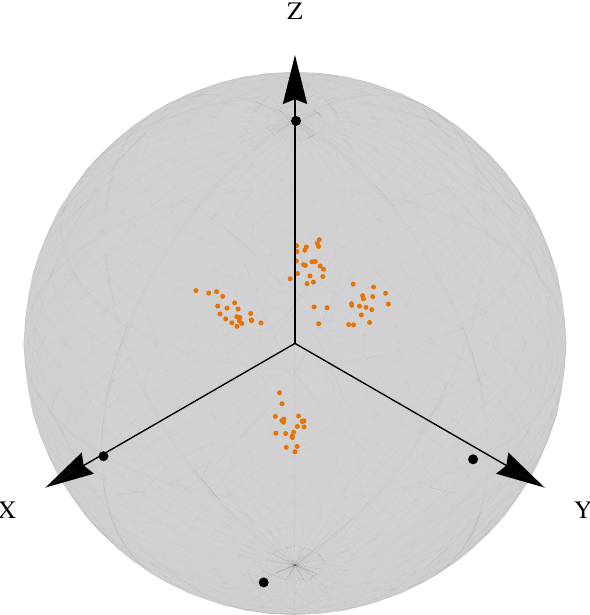}
\end{tabular}
\caption{Experimentally reconstructed input and output states of erroneous maps $O^{err}_{3,4,6,8}$ when $\phi_e$ = 
5$^\circ$.}
\label{errors}
\end{figure}

Figure \ref{errors} represents the input and output states of erroneous maps $O^{err}_{3,4,6,8}$ for $\phi_e$ = 5$^\circ$. 
The output states of $O_N^{err}$ are more diffused than those of $O_M^{err}$, where $N<M$, and we see that for 
larger $N$, the map $O_N$ is less sensitive to random operational errors. % These results are in good agreement with the predictions in Ref.~\cite{Bang}.
Since the output states have a distribution in the Bloch sphere, we describe the 
erroneous maps $O_N^{err}$ as the average value of $F_N$ and $\Delta_N$ over 20 QPT results.

\begin{figure}[t]
\centerline{\includegraphics[scale=0.32]{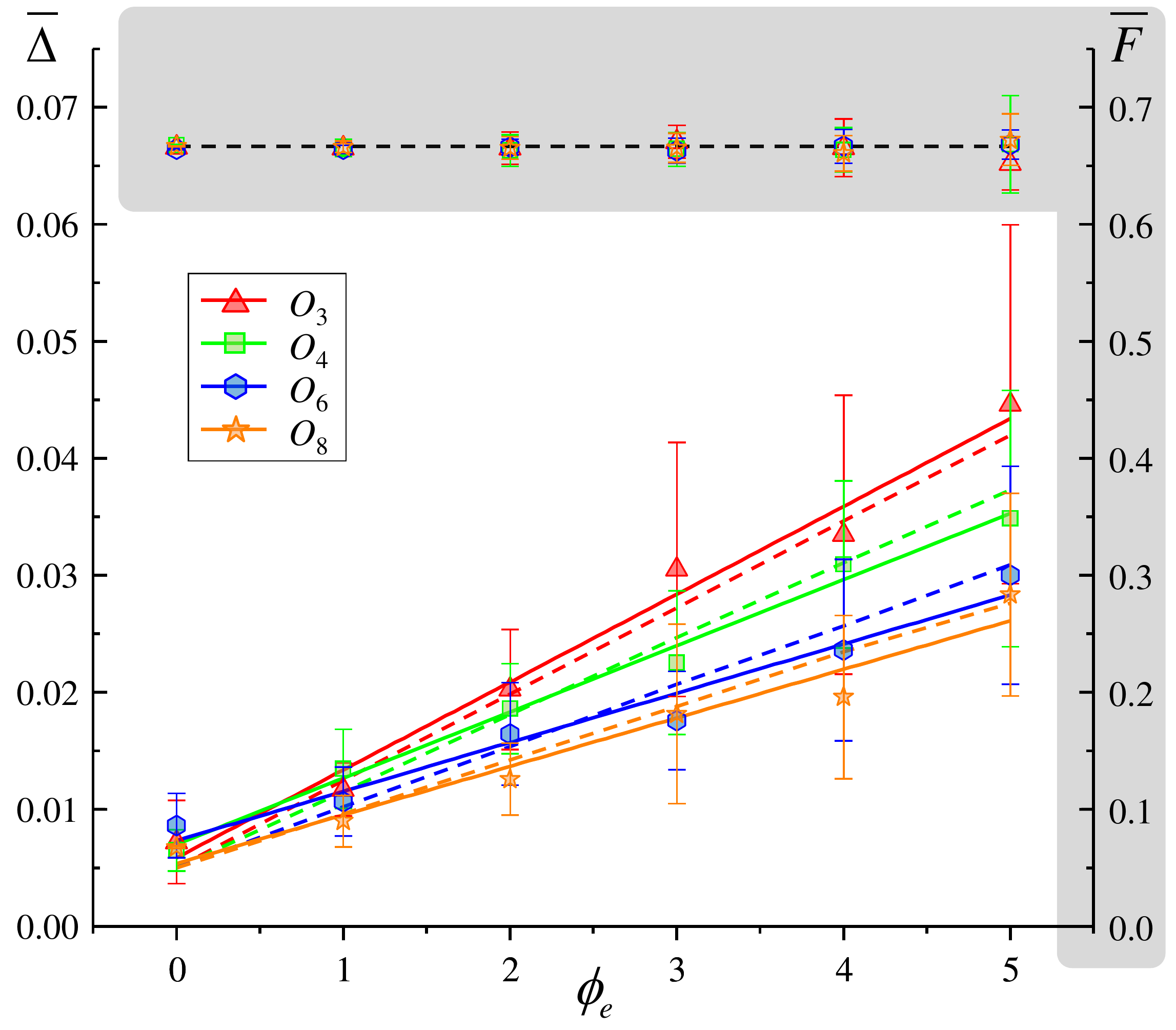}}
\caption{Average values of $F_N$ and $\Delta_N$ for $O^{err}_{3,4,6,8}$ for various random error boundaries $\phi_e$. The 
points and solid lines are the averages of the experimental results and a linear fit of the data. The dashed lines are 
shifted trends of the simulation results for 10$^4$ QPT measurements.}
\label{avfd}
\end{figure}

We repeated the QPT experiments for the maps $O^{err}_{3,4,6,8}$ for various error boundaries $\phi_e$ to measure 
$\mathcal{S}_N$. The average values of $F_N$ and $\Delta_N$ for random errors are shown in Fig.~\ref{avfd}. Each point
represents the average value of 20 experimental results, except for the error-free case ($\phi_e$ = 0$^\circ$) in which 2 results were averaged. The solid and dashed lines represent linear fits of the experimental data and trends of the simulation 
results for $10^4$ QPT measurements, respectively. The simulation results are shifted up about 0.005 to match the 
experimental results. We attribute this offset to inevitable imperfections in the experimental setup and measurements.

%figures

In Fig.~\ref{avfd}, the results of $\overline{\Delta}_N$ are distinct for different $N$, although there are no significant 
differences in $\overline{F}_N$. The experimental data of $\overline{\Delta}_N$ and $\overline{F}_N$ agree well with the 
simulation results to within the deviations. From Eq.~(\ref{dev}), we note that the gradients of the $\overline{\Delta}_N$ 
curves corresponding to the sensitivity $\mathcal{S}_N$, and the relative gradients $\mathcal{S}_N/\mathcal{S}_M$ obtained  
from the experimental data (simulation results) for $\overline{\Delta}_N$ and $\overline{\Delta}_M$ obey the relation 
$\sqrt{M/N}$ to within an accuracy of about 82$\%$ (99$\%$). This shows that the sensitivity of the stochastic map $O_N$ is 
inversely proportional to the square root of the number of operations $N$ as described in Eq.~(\ref{dev}).

Note that all the experimental data (1 QST or 1 QPT) were measured using the same amount of resources (photon pairs). Thus, the error insensitivity of $O_N$ 
is not a matter of the measurement precision dependent on the number of repetitions. 
The essential point of our scheme is that the output state is a convex combination of other states, so that we expect that our error-insensitive method will be applicable to other stochastic operations. 
We utilized the stochastic method 
rather than the ancillar-assisted model, which demands $N - 1$ controlled unitary operations for $O_N$, so as to 
avoid any controlled unitary operations which are probabilistic in the framework of linear optics. 
Most cases of the ancillar-assisted model are composed of sequential controlled unitary operations so that
their operational errors can be accumulated. Therefore, to apply our method to the ancillar-assisted model, a specific 
controlled operation and an equally superposed ancillar state are required, e.g., 
$U_{AB}^{(N)}=\sum_{i=1}^N|i\rangle_A\langle i| \otimes (\vec \sigma \cdot \vec n _i )_B$ and $|\psi_0\rangle_A = \frac 1 
{\sqrt{N}} \sum_{i=1}^N|i\rangle$, where $A$ and $B$ denote ancillar and object systems, respectively.

%section
{\it Conclusion.-}
We have introduced an error-insensitive (robust) U-NOT gate consisting of stochastic unitary operations with 
rotation axes corresponding to the vertex directions of an octahedron and hexahedron and a rotation angle of $\pi$. 
We demonstrated both theoretically and experimentally that the sensitivity of the map to random operational errors is 
inversely proportional to the square root of the number of stochastic operations. The method does not require any 
increase in the total number of measurements nor additional resources. Even though we have considered only the maps $O_N$ for 
$N$ = 3, 4, 6, and 8, our scheme can be generalized to $N$ = $3n$ and $4n$ (1 $\leq$ $n$). This method is also applicable to 
all approximate anti-unitary operations, since such operations are equivalent to a unitary transformation, and it may be 
possible to extend the method to other stochastic mappings.

%acknowledgments
S.M.L. and H.S.M. thank to Hee Su Park for fruitful discussions.
This research was supported by the Basic Science Research Program through the National Research Foundation of Korea (NRF) funded by the Ministry of Education, Science and Technology (Grant \#2012R1A2A1A01006579,  2013R1A4A1069587, 2010-0015059, and 2010-0018295).

%\bibliographystyle{apsrev4-1}
%\bibliography{UNOT01}

\begin{thebibliography}{25}
\bibitem [{pol()}]{poly}%
  \BibitemOpen
  \href@noop {} {}\bibinfo {note} {The number of vertices for tetra-, hexa-,
  octa-, dodeca-, and icosa-hedron are 4, 8, 6, 20, and 12, respectively.
  However, the half of points for each polyhedron are located on opposite sites
  of the other half, except tetrahedron. Thus, the minimum number of stochastic
  operations, to satisfy the universality condition, represented by each
  poly-hodrons is 4, 4, 3, 10, and 6, respectively. The cases of tetra- and
  hexa-hedron are equivalent.}\BibitemShut {Stop}%
\bibitem [{f2()}]{f2}%
  \BibitemOpen
  \href@noop {} {}\bibinfo {note} {The $\vec e _i$ has symmetric and random
  distribution on an orthogonal plane to $\vec o _i$ when $| \vec e _i | \ll
  1$}\BibitemShut {NoStop}%
\bibitem [{EM()}]{EM}%
  \BibitemOpen
  \href@noop {} {}\bibinfo {note} {Although experimental uncertainty of optical
  axis is less than 2$^\circ$ in general, we examine up to the case of
  $\phi_e$=5$^\circ$ to clearly show the $\mathcal{S}_N$. The standard
  deviation of random errors between [$-\phi_e$,$\phi_e$] is proportional to
  $\phi_e$.}\BibitemShut {Stop}%
\bibitem [{pcf()}]{pcf}%
  \BibitemOpen
  \href@noop {} {}\bibinfo {note} {We repeated QSTs of all the input and output
  states for 10 times to analyze effects of photon counting fluctuations.
  However, the distribution of reconstructed states is less than the size of a
  point on the Bloch sphere. So we can neglect photon counting fluctuation
  errors.}\BibitemShut {Stop}%
\end{thebibliography}


\newpage

~

\newpage

\appendix
\section{Supplementary material}

\subsection{Calculation of $F$ and $\Delta$ from $\chi$-matrix}  
When a quantum operation $O$ is charaterized by a $\chi$-matrix as follows
\begin{align}
\chi=
\left( {\begin{array}{*{20}{c}}
  {{\chi _{11}}}&{{\chi _{12}}}&{{\chi _{13}}}&{{\chi _{14}}} \\ 
  {\chi _{12} ^ * }&{{\chi _{22}}}&{{\chi _{23}}}&{{\chi _{24}}} \\ 
  {\chi _{13} ^ * }&{\chi _{23} ^ * }&{{\chi _{33}}}&{{\chi _{34}}} \\ 
  {\chi _{14} ^ * }&{\chi _{24} ^ * }&{ \chi _{34} ^ * }&{{\chi _{44}}} 
\end{array}} \right),
\end{align}
the output state is expressed as 
\begin{align}
O(\psi)=\sum_{i,j=0} ^3 \chi_{ij} ~\sigma_i |\psi\rangle \langle \psi| \sigma_j ,
\end{align}
where $\sigma_0$ is $2 \times 2$ identiy matrix and $\sigma_{i\neq0}$ are the Pauli matrices. The fidelity between the output state of the map $O$ and ideal U-NOT gate is described as
\begin{align}
f(O(\psi),\psi_\bot)
&= \langle \psi_\bot| O(\psi) | \psi_\bot \rangle \notag\\
&=\sum_{i,j=0} ^3 \chi_{ij}  \langle \psi_\bot| \sigma_i |\psi\rangle \langle \psi| \sigma_j | \psi_\bot \rangle \notag \\
&=\sum_{i,j=0} ^3 \chi_{ij}  C_i C_j^*,
\end{align}
where the coefficients are defined as $C_i$$\equiv$$\langle \psi_\bot| \sigma_i |\psi\rangle$. Since an arbitrary pure state and its orthogonal state can be represented by $|\psi\rangle$=$\cos\frac{\theta}{2}|0\rangle + e^{i \phi} \sin\frac{\theta}{2}|1\rangle$ and $|\psi_\bot\rangle$=$\sin\frac{\theta}{2}|0\rangle - e^{i \phi} \cos\frac{\theta}{2}|1\rangle$, respectively, where $\theta$=$[0,\pi]$ and $\phi$=$[0,2\pi]$, the coefficients $C_i$ are functions of $\theta$ and $\phi$. Thus, the average fideilty $F$ and the square of the fidelity deviation $\Delta^2$ are obtained as
\begin{align}
F 
& = \frac{1}{4\pi} \sum_{i,j=0} ^3 \chi_{ij} \iint  C_i C_j^* \sin \theta ~ d \theta d\phi  \notag \\
& =  \frac{2}{3}\left( {{\chi _{11}} + {\chi _{22}} + {\chi _{33}}} \right), \label{fid_cal} \\
\Delta^2 
& = \frac{1}{4\pi} \sum_{i,j,k,l=0} ^3 \chi_{ij} \chi_{kl}  \iint   C_i C_j^* C_k C_l^* \sin \theta ~ d \theta d\phi - F^2 \notag \\
& = \frac{4}{45} \left( \chi_{11} ^2 + \chi_{22} ^2 + \chi_{33} ^2  - \chi _{11} \chi _{22} - \chi _{11} \chi _{33} - \chi _{22} \chi _{33} \right) \notag \\
& ~~+ \frac{4}{15} \left( 3 \left| \chi_{12} \right| ^2  + 3 \left| \chi_{13} \right| ^2 + 3 \left| \chi_{23} \right| ^2 \right.  \notag \\
& ~~~~~~~~~~~~\left.  - 2Re \left[ \chi_{12} ^2 + \chi_{13} ^2 + \chi _{23} ^2 \right] \right). \label{dev_cal}
\end{align}

\subsection{Random operational error and $O_N^{err}$}
A random operational error can be considered as an additional random unitary operation following the original operation: $\vec \sigma \cdot \vec n _i$$\xrightarrow{error}$$V_i (\vec \sigma \cdot \vec n _i)$. The error operation is defined as $V_i= e^{i {\vec \epsilon _i} \cdot \vec \sigma} \simeq  I +  i ~{\vec \epsilon _i} \cdot \vec \sigma$ where $\vec \epsilon _i$=$(\epsilon _{i1},\epsilon _{i2},\epsilon _{i3})$ and $ |\epsilon _{ij}| \leq \epsilon_0 \ll 1$. The distribution of errors, $P(\vec \epsilon _i)$ is assumed to be homogeneous and symmetric under the inversion: $P(\vec \epsilon _i) = P(-\vec \epsilon _i)$ so that $\overline{\vec \epsilon _i} = 0$. Then, the erroneous map $O_N^{err}$ is expressed as
\begin{gather}
\rho \rightarrow \rho_N''=O_N^{err}(\rho)=\frac{1}{N}\sum_{i=1}^{N} V_{i} (\vec \sigma \cdot \vec n _i) \rho  (\vec \sigma \cdot \vec n _i) V_i ^\dagger.
\end{gather}

\subsection{Proof of Eq.~(5)}
\subsubsection{The case of $O_3^{err}$} 
The first order of error terms $\delta O_3^{(1)}$ for the map $O_3^{err}$ and their additional contribution $\delta \chi_3 ^{(1)}$ to $\chi_I$-matrix are as follows,
\begin{align}
\delta O_3^{(1)}=&
\frac{1}{3}\left[i({\vec \epsilon _1} \cdot \vec \sigma) \sigma_x \rho \sigma_x+ i( {\vec \epsilon _2} \cdot \vec \sigma) \sigma_y \rho \sigma_y \right. \notag\\
& ~ ~ ~ \left. + i( {\vec \epsilon _3} \cdot \vec \sigma) \sigma_z \rho \sigma_z + c.c.\right],
\\
\delta \chi_3 ^{(1)}=& 
\frac{1}{3} 
\left( {\begin{array}{*{20}{c}}
  0&{i{\epsilon _{11}}}&{i{\epsilon _{22}}}&{i{\epsilon _{33}}} \\ 
  { - i{\epsilon _{11}}}&0&{{\epsilon _{23}} - {\epsilon _{13}}}&{{\epsilon _{12}} - {\epsilon _{32}}} \\ 
  { - i{\epsilon _{22}}}&{{\epsilon _{23}} - {\epsilon _{13}}}&0&{{\epsilon _{31}} - {\epsilon _{21}}} \\ 
  { - i{\epsilon _{33}}}&{{\epsilon _{12}} - {\epsilon _{32}}}&{{\epsilon _{31}} - {\epsilon _{21}}}&0 
\label{chi3_err}  
\end{array}} \right).
\end{align}
Using Eqs.~(\ref{fid_cal}), (\ref{dev_cal}) and (\ref{chi3_err}), the average fidelity and the fidelity deviation for $O_3^{err}$ are obtained as
\begin{align}
F_{3}&=2/3 , \label{fid_3} \\
\Delta_{3}
&=\frac{2 \sqrt{ (\epsilon_{13}-\epsilon_{23})^2 +(\epsilon_{21}-\epsilon_{31})^2  +(\epsilon_{12}-\epsilon_{32})^2 } }{3\sqrt{15}} \notag \\
&=\frac{2\sqrt{2}}{3\sqrt{15}} \sqrt{  r_1 ^2 + r_2 ^2 +r_3 ^2 },
\label{dev_3}
\end{align}
where $\{r_i\}$ are replaced random variables of $\{ \epsilon_{ij} \}$ up to normalization factor $\sqrt{2}$. Note that the errors of which directions are parallel to the original operations, i.e., $\{ \epsilon_{ii} \}$, do not contribute to the average fidelity and the fidelity deviation as shown in Eqs.~(\ref{chi3_err}), (\ref{fid_3}) and (\ref{dev_3}). A mean of the fidelity deviation over random errors is propotional to standard deviation of random variables,
\begin{gather}
\overline{\Delta}_{3}=\sqrt{\frac{8 }{15} } ~ \frac{\delta_r}{\sqrt{  3}}.
\label{ad3}
\end{gather}
where  $\delta_r = \overline{ r_i ^2} ^{\frac 1 2}$, and we assume $\overline{ r_i}=0$.

\subsubsection{The case of $O_4^{err}$}
After tedious calculations, the first order of error terms $\delta O_4^{(1)}$ and their contribution $\delta\chi_4^{(1)}$ to $\chi_I$-matrix are expressed as
\begin{widetext}
\begin{align}
\delta O_4^{(1)}=
\frac{-1}{12} & \left[
~ \left(\epsilon_{11} \sqrt{3} \sigma_z +  \epsilon_{12}  ( \sigma_x - \sqrt{2} \sigma_y)\right) \rho ( \sqrt{2} \sigma_x + \sigma_y) 
+ \left(\epsilon_{21} \sqrt{3} \sigma_z +  \epsilon_{22}  ( \sigma_x + \sqrt{2} \sigma_y)\right) \rho (-\sqrt{2} \sigma_x + \sigma_y) \right. \\ 
& ~ \left.
+ \left(\epsilon_{31} \sqrt{3} \sigma_x +  \epsilon_{32}  ( \sqrt{2} \sigma_y + \sigma_z)\right) \rho (-\sigma_y + \sqrt{2} \sigma_z) 
+ \left(\epsilon_{41} \sqrt{3} \sigma_x +  \epsilon_{42}  (-\sqrt{2} \sigma_y + \sigma_z)\right) \rho (-\sigma_y - \sqrt{2} \sigma_z)
 +c.c. ~ \right], \notag
\end{align}
\begin{gather}
\delta\chi_4^{(1)}=\frac{1}{{12}}\left( {\begin{array}{*{20}{c}}
  0&0&0&0 \\ 
  0&{2\sqrt 2 ( - {\epsilon _{12}} + {\epsilon _{22}})}&{{\epsilon _{12}} + {\epsilon _{22}} + \sqrt 3 ({\epsilon _{31}} + {\epsilon _{41}})}&{\sqrt 6 ( - {\epsilon _{11}} + {\epsilon _{21}} - {\epsilon _{31}} + {\epsilon _{41}})} \\ 
  0&{{\epsilon _{12}} + {\epsilon _{22}} + \sqrt 3 ({\epsilon _{31}} + {\epsilon _{41}})}&{2\sqrt 2 ({\epsilon _{12}} - {\epsilon _{22}} + {\epsilon _{32}} - {\epsilon _{42}})}&{ - \sqrt 3 ({\epsilon _{11}} + {\epsilon _{21}}) - {\epsilon _{32}} - {\epsilon _{42}}} \\ 
  0&{\sqrt 6 ( - {\epsilon _{11}} + {\epsilon _{21}} - {\epsilon _{31}} + {\epsilon _{41}})}&{ - \sqrt 3 ({\epsilon _{11}} + {\epsilon _{21}}) - {\epsilon _{32}} - {\epsilon _{42}}}&{2\sqrt 2 ( - {\epsilon _{32}} + {\epsilon _{42}})} 
\label{chi4_err}
\end{array}} \right).
\end{gather}
In this case, for a simple calculation, we neglect errors which are parallel to the original operations, since the errors do not contribute to the first order calculation as shown in the case of $O_3^{err}$. Using Eqs.~(\ref{fid_cal}), (\ref{dev_cal}), (\ref{chi4_err}) and replaced random variables, $F_{4}$ and $\Delta_{4}$ for $O_4^{err}$ are obtained as
\begin{align}
F_{4}&=2/3, \\
\Delta_{4}
&=\frac{1}{3\sqrt{30}} \sqrt{ ( \sqrt{3} \alpha_+ + \delta_+  )^2 + 6 ( \alpha_- + \gamma_-  )^2 +  ( \beta_+ + \sqrt{3} \gamma_+  )^2 + 8 (\beta_-^2  + \beta_- \delta_- + \delta_-^2 ) } \notag \\
&=\frac{\sqrt{2}}{3\sqrt{15}} \sqrt{ R_1^2  + 3 R_2^2 + R_3^2 +3 R_4^2 + R_5^2   } ,
\end{align}
where the replaced variables are defined as
\begin{gather}
\alpha_\pm =  \frac{\epsilon_{11} \pm  \epsilon_{21} }{\sqrt{2}} , ~~~
\beta_\pm  =  \frac{\epsilon_{12} \pm  \epsilon_{22} }{\sqrt{2}} , ~~~
\gamma_\pm =  \frac{\epsilon_{31} \pm  \epsilon_{41} }{\sqrt{2}} , ~~~
\delta_\pm =  \frac{\epsilon_{32} \pm  \epsilon_{42} }{\sqrt{2}} , \\
R_1=\frac{ \sqrt{3} \alpha_+ + \delta_+ }{2}, ~~~
R_2=\frac{ \alpha_- + \gamma_- }{\sqrt{2}}, ~~~
R_3=\frac{ \beta_+ + \sqrt{3} \gamma_+ }{2}, ~~~
R_4=\frac{ \beta_- + \delta_- }{\sqrt{2}}, ~~~
R_5=\frac{ \beta_- - \delta_- }{\sqrt{2}}.
\end{gather}
\end{widetext}
A mean of $\Delta_{4}$ over random errors is obtained as
\begin{gather}
\overline{\Delta}_{4}=\sqrt{\frac{8 }{15} } ~  \frac{\delta_r}{\sqrt{4} }.
\label{ad4}
\end{gather}

\subsubsection{The cases of $O_6^{err}$ and $O_8^{err}$} 
From tedious calculations, average fidelities are the same as 2/3, and means of $\Delta_{6}$ and $\Delta_{6}$ over random errors are obtained as
\begin{gather}
\overline{\Delta}_{6}=\sqrt{\frac{8 }{15} } ~ \frac{\delta_r}{\sqrt{6} }, \label{ad6} \\
\overline{\Delta}_{8}=\sqrt{\frac{8 }{15} } ~ \frac{\delta_r}{\sqrt{8} }. \label{ad8}
\end{gather}
From Eqs.~(\ref{ad3}), (\ref{ad4}), (\ref{ad6}) and (\ref{ad8}), we infer that the average of $\Delta_N$ (at least for the cases of $N$=3$n$, 4$n$) over random errors is expressed as
\begin{gather}
\overline{\Delta}_{N}=\sqrt{\frac{8 }{15} } ~ \frac{\delta_r}{\sqrt{N} }.
\label{general}
\end{gather}

\end{document}